\documentclass[reprint,aps,prb,groupedaddress,showpacs]{revtex4-1}
\usepackage{graphicx}
\usepackage{color}
\usepackage{dcolumn}
\usepackage{bm}
\usepackage{threeparttable}
\usepackage{rotfloat}
\usepackage{float}
\usepackage{amsmath}
\def\degree{${}^{\circ}$}

\begin{document}
\title{%
Strain-induced quantum topological phase transitions in Na$_3$Bi
}

\author{Dexi Shao$^1$}
\author{Jiawei Ruan$^1$}
\author{Juefei Wu$^1$}
\author{Tong Chen$^1$}
\author{Zhaopeng Guo$^1$}
\author{Haijun Zhang$^{1,2}$}
\author{Jian Sun$^{1,2}$}
\email[]{To whom correspondence should be addressed. E-mail: jiansun@nju.edu.cn}

\author{Li Sheng$^{1,2}$}
\author{Dingyu Xing$^{1,2}$}

\affiliation{
$^1$National Laboratory of Solid State Microstructures and
Department of Physics, Nanjing University, Nanjing 210093, China\\
$^2$ Collaborative Innovation Center of Advanced Microstructures, Nanjing 210093, China}

\begin{abstract}

Strain can be used as an effective tool
to tune the crystal structure of materials and hence to
modify their electronic structures, including topological properties.
Here, taking Na$_3$Bi as a paradigmatic example,
we demonstrated with first-principles calculations
and $\mathbf{k}\cdot\mathbf{p}$ models
that the topological phase transitions can be induced
by various types of strains.
For instance,
the Dirac semimetal phase of ambient Na$_3$Bi can be
tuned into a topological insulator (TI) phase
by uniaxial strain along the $\langle100\rangle$ axis.
Hydrostatic pressure can let the ambient structure transfer into
a new thermodynamically stable phase with Fm$\bar{3}$m symmetry,
coming with a perfect parabolic semimetal
having a single contact point between the conduction and valence bands,
exactly at $\Gamma$ point on the Fermi level like $\alpha$-Sn.
Furthermore, uniaxial strain in the $\langle100\rangle$ direction can tune
the new parabolic semimetal phase into a Dirac semimetal,
while shear strains in both the $\langle100\rangle$ and
$\langle111\rangle$ directions can take the new parabolic semimetal phase into a TI.
$\mathbf{k}\cdot\mathbf{p}$ models
are constructed to gain more insights
into these quantum topological phase transitions.
At last, we calculated surface states of Fm$\bar{3}$m Na$_3$Bi without and with strains
to verify these topological transitions.

\end{abstract}

\maketitle

\section{INTRODUCTION}
Due to the inspiration from fundamental physics and interest in exotic properties for applications,
new topological materials and phases with non-trivial band topology,
such as topological insulators (TIs), topological metals/semimetals and
topological superconductors,
attracted tremendous attention in the past decade.
~\cite{TIs-Rmp-2010-Hasan,TIs-with-inversion-2011-(1057)FuLiang,TSM-Burkov-2011,WengHM2016-JPCM-SM-review}
The non-trivial states are usually protected by certain symmetries,
such as time-reversal symmetry (TRS), crystalline symmetry including
inversion symmetry (IS).
Among these topological states,
TRS-protected surface states
were first predicted in 1987~\cite{Pb-xSnxTe-Hg-xCdxTe-O.A.Pankratov-1987}
to occur in quantum wells of HgTe sandwiched between CdTe
and were successfully observed in experiments in 2006~\cite{HgTe-ZSC}.
In 3D TIs ,
the surface state is actually a new type of two-dimensional (2D) massless electron gas,
with its spin locked to its momentum
~\cite{Tunable-TIs-S-helical-Dirac-reg-Hsieh-2009,Bi2X3-Dirac-surface-NP2009-ZHJ}.
These robust metallic surface states differ TIs from normal insulators
and make TIs greatly attractive.

Compared with TIs indexed by Z$_{2}$,
topological semimetals, in which band crossings
appear at Fermi level in a reduced dimension,
have attracted more attention because they may supply platforms
to investigate new types of fermion-like excitations,
including Dirac fermions~\cite{DS-AB3-Wang-2012,3DDS-Cd3As2-Wang-2013},
Weyl fermions~\cite{Wan2011-weyl,HgCrSe-2011,Weng2015-TaAs,Xu2015-TaAs}
and nodal lines~\cite{graphene-network,Cu3PdN,Cu3N-newZ2,Bian2016-PbTaSe2,Bzdu2016Nodal,nodalline-alkali-2016}, etc.
Among them,
the earliest example may be the 2D Dirac semimetal --
graphene~\cite{2d-graphene-DF-Novoselov-2005,Rise-of-graphene-Geim-2007}.
Interestingly, later works show that the surface states in 3D TIs
also present 2D massless Dirac-like dispersions.
Among many candidates with Dirac-like dispersions have been
reported~\cite{DS-AB3-Wang-2012,3DDS-Cd3As2-Wang-2013,3D-BiO2,3dssm-design},
Cd$_{3}$As$_{2}$ and Na$_{3}$Bi are particularly attractive
3D Dirac semimetals with their Dirac points locating exactly at the Fermi level.
Featured by 3D Dirac points in the bulk and Fermi arcs
on the surface~\cite{DS-AB3-Wang-2012,3DDS-Cd3As2-Wang-2013,3DTIs-Hasan2010,3D-BiO2},
3D Dirac semimetals have recently been identified experimentally
in Cd$_{3}$As$_{2}$~\cite{Cd3As2-Liu-NM-2014,Observation-Cd3As2-NC-XuSuYang-2014,
Exp-realization-DS-Cava-2014,Cd3As2:Evi-surface-state-Yi-2013,Landau-quan-CdAs-NM-Jeon-2014}
and Na$_{3}$Bi systems~\cite{Exp-verify-Na3Bi-Sci-2014,Exp-fermi-arc-Sci-XSY-2015}.

Na$_{3}$Bi was predicted to be a 3D bulk Dirac semimetal (DSM)~\cite{DS-AB3-Wang-2012}
and verified later by experiments~\cite{Exp-verify-Na3Bi-Sci-2014}.
This topological Dirac fermion in Na$_{3}$Bi is
protected by TRS and IS together with R$_{3z}$ symmetry.
It is known that breaking of the R$_{3z}$ symmetry,
for example 1\% compression along the y axis,
will change the system into a TI with
$Z_{2}=1$~\cite{TIs-with-inversion-2011-(1057)FuLiang, TIs-Rmp-2010-Hasan}.
After the predictions of Na$_{3}$Bi to be a 3D Dirac semimetal,
large efforts were invested to study this interesting system,
for instance, Fermi arc surface states~\cite{Exp-fermi-arc-Sci-XSY-2015},
quantum oscillations~\cite{Quantum2014Potter},
evidence for the chiral anomaly~\cite{chiral-anomaly-sci-Xiong-2015},
magnetoresistance~\cite{PhysRevB.92.075205}, etc.
As we know, alkali pnictides A$_{3}$B (A = alkali metal, B = pnictide)
usually crystalline into two different structures at ambient pressure:
the hexagonal P$6_{3}$/mmc phase (e.g., K$_{3}$Bi and Rb$_{3}$Bi)
and the cubic Fm$\bar{3}$m phase (e.g., Li$_{3}$Bi and Cs$_{3}$Bi)~\cite{High-pressure-Syn-Kulinich-1999}.
Actually, phonon spectra of P$6_{3}$/mmc phase of Na$_{3}$Bi show negative frequencies,
which means this phase should not be dynamically stable,
Cheng et al.~\cite{Ground-state-Na3Bi-Cheng-2014} reported that
the ground state of Na$_{3}$Bi at ambient pressure
could be a P$\bar{3}$c1 (or so-called hP24) phase,
which is a distorted superlattice version of the P$6_{3}$/mmc phase.
And this P$\bar{3}$c1 phase also exhibits features of 3D Dirac semimetal.

Pressure and strain have been used as effective methods to
modify the topological properties of materials,
for instance, in graphene~\cite{Guinea2009-graphene},
BiTeI~\cite{Bahramy-BiTeI-pressure-2012},
HgTe-class\cite{WeylSM-HgTe-class-Ruan-NC-2016},
Cd$_{3}$As$_{2}$~\cite{Cd3As2SLL,Cd3As2-LSy,Cd3As2-strain-SC}, TaAs~\cite{TaAs-ours},
ZrTe$_{5}$~\cite{ZrTe5}, WTe$_{2}$~\cite{WTe2-Lu},
SnTe\cite{Chen2017-SnTe-SC}, TaP~\cite{TaP-SC}, etc.
The work by Cheng et al.~\cite{NaBi-system-2015, deform-Fm-3m-Na3Bi-Cheng-2015}
showed that Na$_{3}$Bi would undergo a pressure-induced structural phase transition
from the P$\bar{3}$c1 (hP24) phase to a cubic Fm$\bar{3}$m (cF16) phase at pressure of about 0.8 GPa,
which is in good agreement with previous experimental
findings~\cite{exp-Na3Bi-1998,HPre-Hex-Alkali-Pnictides-2003,High-pressure-Syn-Kulinich-1999}.
The transition pressure for Na$_{3}$Bi is such low,
which means compressive strain has a large effect
on the structure of this system.
Previous work~\cite{deform-Fm-3m-Na3Bi-Cheng-2015} reported that
shear strain along $\langle100\rangle$ axis
can develop the cubic phase of Na$_{3}$Bi into a TI.
However, whether or how different type of strain will affect the topological properties
of Na$_{3}$Bi and detailed analysis with model Hamiltonian still remains an open question.

In this work, we have studied effects of several different strains on Na$_{3}$Bi,
including uniaxial tensile/compressive strain
and shear strain in different directions.
We find that uniaxial strain on Na$_{3}$Bi
in the space group of P$\bar{3}$c1 at ambient pressure
can induce a topological phase transition from Dirac semimetal to TI.
With the help of Luttinger Hamiltonian~\cite{LuttingerH-1956},
we also find the Fm$\overline{3}$m phase to be a perfect parabolic semimetal.
We then impose a uniaxial strain on the
Fm$\overline{3}$m lattice
and find it will
open a gap at the $\Gamma$ point and induce a Dirac crossing
near the $\Gamma$.
Furthermore, with the help of a $\mathbf{k}\cdot\mathbf{p}$ model,
we confirm that the term of $g\Gamma^{5}$ in the Hamiltonian generated from the crystal-field splitting
induces the gap and the Dirac crossing in the uniaxial strained structure.

On the other hand, shear strain in the $\langle100\rangle$ and
$\langle111\rangle$ directions can tune the Fm$\overline{3}$m phase into a TI.
To get more insights of these topological phase transitions,
we develop two $\mathbf{k}\cdot\mathbf{p}$ models
corresponding to the two different shear strains, respectively, and find
that spin-orbit coupling (SOC) together with the splitting of
the crystal field plays a key role in these transitions.

\section{METHODOLOGY AND THE DETAIL}
\emph{Ab initio} random structure searching \cite{Pickard2006-SiH4,Pickard2011-airss-review}
is applied for crystal structure searching under pressure.
Structure optimization is performed using
projector augmented wave (PAW) potential~\cite{PAW}
with the Perdew-Burke-Erzernhof~\cite{PBE}
generalized gradient approximation (PBE-GGA) exchange-correlation functional
implemented in the Vienna \emph{ab initio} simulation package (VASP)~\cite{VASP-tot-energy-Kresse-1996}
in the framework of density functional theory (DFT).
The plane wave cutoff is set to 850 eV, structure relaxation
is carried out until all of the atomic forces on each
ion is less than 0.0025 eV/${\AA}$.
Electronic band structures calculations
are carried out using full-potential linearized
augmented plane-wave method implemented in the WIEN2k~\cite{wien2k} package.
SOC is taken into account self-consistently.
$21\times21\times21$ k-mesh is used as the period boundary condition
for electronic structure calculation under ambient and high pressure.

\begin{figure}[th]
\begin{center}
\includegraphics[width=0.48\textwidth]{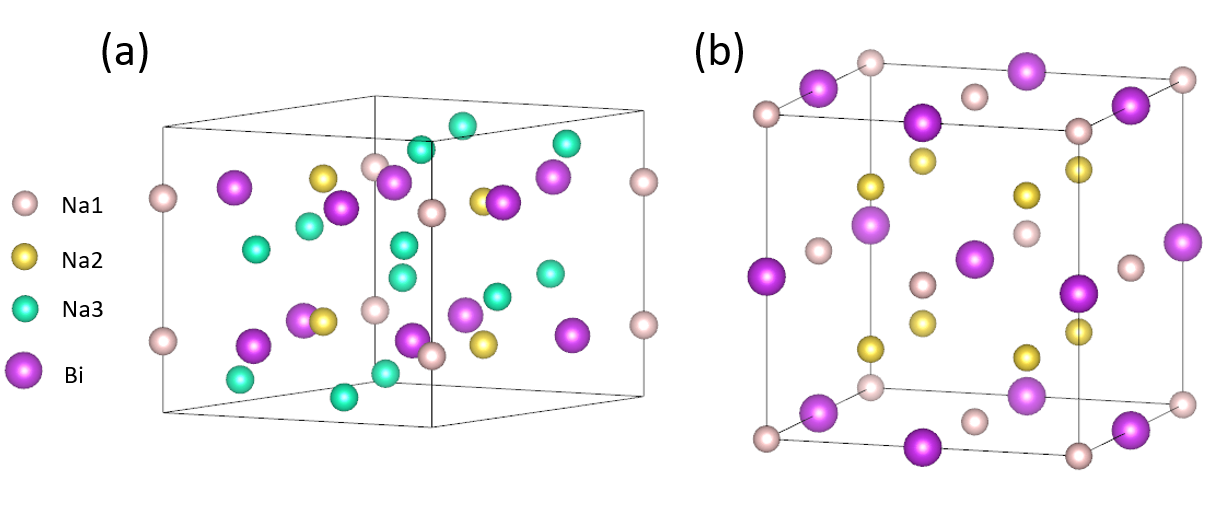}
\caption{%
  (a) Crystal structure of Na$_{3}$Bi at the ambient pressure with P$\bar{3}$c1 symmetry.
Na1, Na2 and Na3 atoms occupy the 2a(0,0,$\frac{1}{4}$),
4d($\frac{1}{3}$,$\frac{2}{3}$,0.200) and 12g(0.354,0.319,0.083) sites, respectively,
while Bi atoms lie at the 6f(0.337,0,$\frac{1}{4}$) site.
(b) Crystal structure of Na$_{3}$Bi at 1 GPa with Fm$\bar{3}$m symmetry.
Na1 and Na2 atoms occupy the 4a(0,0,0), 8c($\frac{3}{4}$,$\frac{1}{4}$,$\frac{1}{4}$) sites
while Bi atoms stay at the 4b(0,0,$\frac{1}{2}$) site.
}
\label{fig:str-p-3c1-fm-3m}
\end{center}
\end{figure}

\section{THE ELECTRONIC STRUCTURES OF P$\bar{3}$c1 PHASE UNDER UNIAXIAL STRAIN}
A recent work reveals that the zero-pressure ground state of Na$_{3}$Bi
should be the P$\bar{3}$c1 phase
which presents the features of Dirac semimetal~\cite{Ground-state-Na3Bi-Cheng-2014}.
The crystal structures of the P$\bar{3}$c1 phase are shown in Fig.~\ref{fig:str-p-3c1-fm-3m},
and the corresponding lattice parameters are showed in Table.~\ref{table:str-parameters}.

\begin{table}
\begin{ruledtabular}
\caption{Lattice parameters of Na$_{3}$Bi at the ambient pressure with P$\bar{3}$c1 symmetry and 1 GPa with Fm$\bar{3}$m symmetry.}
\begin{tabular}{c c c c c c}
          \textbf{phase} &   \textbf{pressure} (GPa) & \textbf{$a=b$} (${\AA}$)  & \textbf{$c$} (${\AA}$)  &  \textbf{$\alpha=\beta$} (\degree) &  \textbf{$\gamma$} (\degree)\\\hline
           P$\overline{3}$c1   &  0                  & 9.459             & 9.674     &   90                &   120 \\
           Fm$\overline{3}$m   &  1                  & 7.458             & 7.458     &   90                &   90 \\
\end{tabular}
\label{table:str-parameters}
\end{ruledtabular}
\end{table}

\begin{figure*}[th]
\begin{center}
\includegraphics[width=0.95\textwidth]{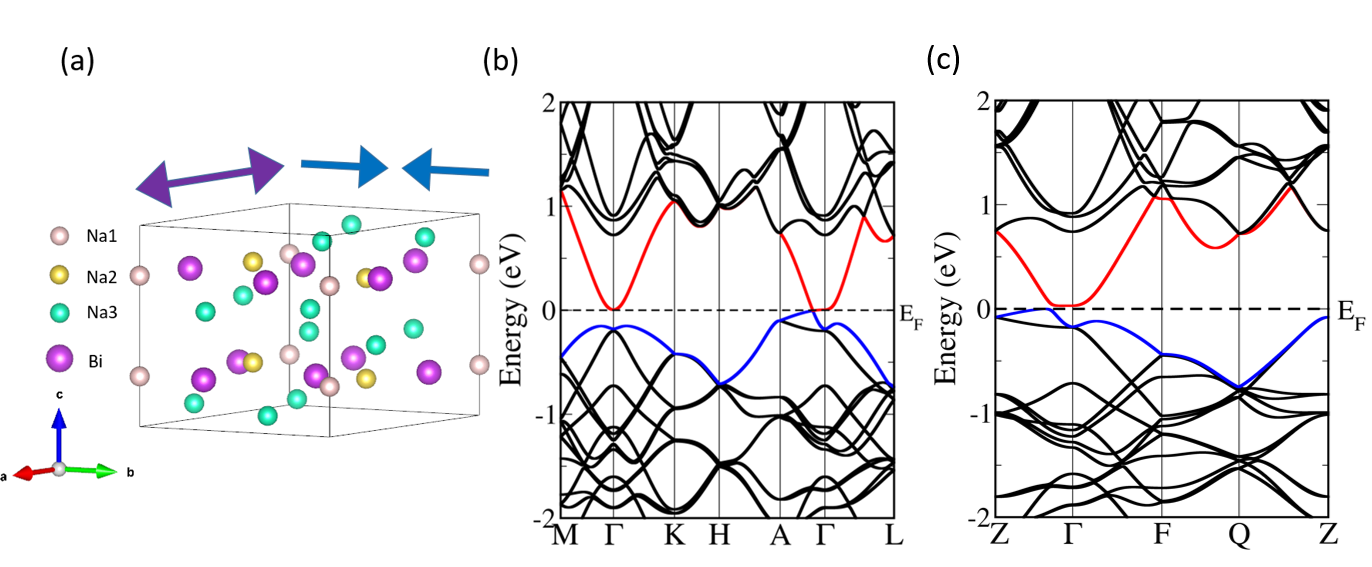}
\caption{%
(a) Crystal structure of Na$_{3}$Bi with P$\bar{3}$c1 symmetry under uniaxial strain.
The purple bidirectional arrow represents tensile strain along x axis,
while the two blue arrows represents compressed strain along y axis.
(b) Corresponding band structure of Na$_{3}$Bi without strain and with SOC.
The red and blue lines correspond to the conduction band minimum (CBM)
and the valence band maximum (VBM), respectively.
(c) Corresponding band structure of Na$_{3}$Bi under uniaxial strain and with SOC.
The red and blue lines correspond to CBM and VBM, respectively.
}
\label{fig:p-3c1-fm-3m-strain}
\end{center}
\end{figure*}

There are 24 atoms in one unit cell of the P$\bar{3}$c1 phase
which occupy 4 nonequivalent positions.
We sign the atoms locate at these nonequivalent positions
with Na1, Na2, Na3 and Bi, respectively,
as shown in Fig.~\ref{fig:str-p-3c1-fm-3m} (a).
The band structure with SOC for the P$\bar{3}$c1 phase Na$_{3}$Bi is shown in Fig.~\ref{fig:p-3c1-fm-3m-strain}(b),
the result is similar to earlier work by Cheng et al.~\cite{Ground-state-Na3Bi-Cheng-2014}
We impose a uniaxial tensile strain along the x axis
and meanwhile a compressed strain along the y axis to keep the volume unchanged.
As shown in Fig.~\ref{fig:p-3c1-fm-3m-strain}(a), this operation or vice versa,
breaks the R$_{3z}$ symmetry in the P$\bar{3}$c1 phase and
changes the space group of the structure from P$\bar{3}$c1 to P$\bar{1}$.
We take 2\% uniaxial strain as an example here,
and the resulting band structure with SOC is shown in Fig.~\ref{fig:p-3c1-fm-3m-strain}(c).
From the band structure we can assert that it is an insulator directly.
Using the method by Fu and Kane~\cite{TIs-with-inversion-2007-FuLiang},
we can easily calculate the $Z_{2}$ index by multiplying all the parities for all
the occupied bands at all time-reversal-invariant momenta (TRIMs).
The results are shown in Table.~\ref{table:parity-P-3c1},
which indicates $Z_{2}=(1,111)$ for this system.
This shows that uniaxial strain will induce a topological phase transition
from Dirac semimetal to TI in the P$\bar{3}$c1 phase.

\begin{table}
\begin{ruledtabular}
\caption{The product of the parities for all the occupied bands
at the eight TRIMs for the P$\bar{3}$c1 phase of Na$_{3}$Bi with uniaxial strain.}
   \begin{tabular}{c c c c c c}%
          \textbf{TRIM}         & $\Gamma$  &  3M  &  3L  &  A    & total\\\hline
           Parity           &    +      &  --  &  --  &  --    & --\\
  \end{tabular}
  \label{table:parity-P-3c1}
\end{ruledtabular}
\end{table}

\section{THE ELECTRONIC STRUCTURES OF THE HIGH-PRESSURE Fm$\bar{3}$m PHASE}
Crystal structure searches and thermodynamic calculations afterwards
show that a cubic phase in the space group of Fm$\bar{3}$m
becomes more stable than the ambient P$\bar{3}$c1 phase
under the critical pressure around 0.8 GPa~\cite{deform-Fm-3m-Na3Bi-Cheng-2015}.

For the Fm$\bar{3}$m phase, there are 4 atoms occupying 3 nonequivalent positions
in one primitive cell as shown in Fig.~\ref{fig:str-p-3c1-fm-3m}(b).
The 4 atoms can be signed as Na1, Na$1^{\prime}$, Na2 and Bi.
Among them, Na1 and Na$1^{\prime}$ are equivalent as a result of IS.

\begin{figure*}[th]
\begin{center}
\includegraphics[width=0.95\textwidth]{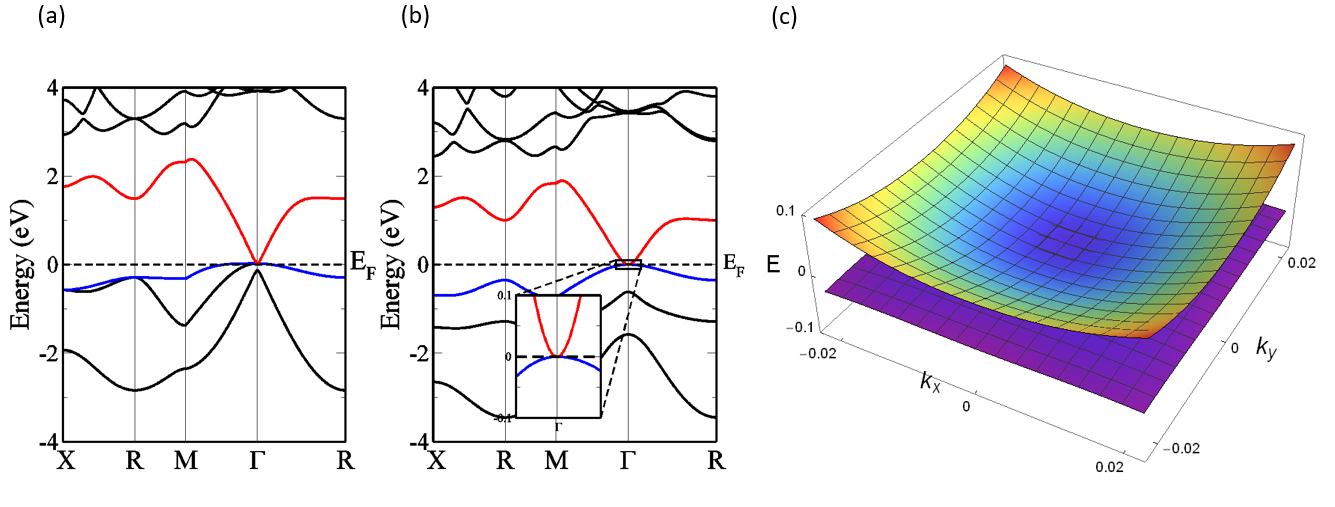}
\caption{%
  The band structure of Fm$\bar{3}$m Na$_{3}$Bi
(a) without soc (b) with soc at 1GPa.
The red and blue lines correspond to CBM and VBM, respectively.
(c) The 2D projected band structure of Fm$\bar{3}$m Na$_{3}$Bi at 1GPa in the plane of kz=0.
}
\label{fig:fm-3m-band}
\end{center}
\end{figure*}

The detailed structural parameters of the Fm$\bar{3}$m phase
are listed in Table.~\ref{table:str-parameters}.
And its electronic band structures without and with SOC
are illustrated in Fig.~\ref{fig:fm-3m-band}(a) and Fig.~\ref{fig:fm-3m-band}(b),
similar to the results by by Cheng et al.~\cite{NaBi-system-2015}

The most interesting feature one can find in the band structures with and without SOC,
as shown in Fig. 3 (a) and (b),
is that there is only one touching point between
the valence and conduction bands,
which exactly locates at $\Gamma$ point on the Fermi level.
Detailed first-principle calculations without SOC indicates that
this touching point is a triply degenerate point contributed most by Bi-6p$_{x,y,z}$ orbits,
and the wave functions of low-energy states around the touching point
mainly consist of Na-3s and Bi-6p$_{x,y,z}$ orbits.
Due the fact that there are two Na1 atoms (signed with Na1 and Na$1^{\prime}$),
which are centrosymmetric connected to each other,
thus we can construct bonding and antibonding states with definite parity
from the s orbitals of Na1 and Na$1^{\prime}$ atoms as follows:
\begin{equation}\label{eq:eps}
|\mathrm{Na1}^{\pm},s>=\frac{1}{\sqrt{2}}(|\mathrm{Na}1;s>\pm|\mathrm{Na}1^{\prime};s>)
\end{equation}
While there is only one atom for Na2 and Bi in the primtive cell,
therefore, the parity of the orbits of Na2 and Bi atoms
are only determined by their orbital angular
quantum number themselves.

Taking SOC into consideration,
spin and orbital angular momentum are coupled together,
which generates a group of new eigenstates with certain total angular quamtum numbers.
We mark these new eigenstates as
$|S_{\mathrm{Na1},\frac{1}{2}}^{\pm},\pm\frac{1}{2}>$,
$|S_{\mathrm{Na2},\frac{1}{2}}^{+},\pm\frac{1}{2}>$,$|S_{\mathrm{Bi},\frac{1}{2}}^{+},\pm\frac{1}{2}>$,
$|P_{\mathrm{Bi},\frac{3}{2}}^{-},\pm\frac{3}{2}>$,$|P_{\mathrm{Bi},\frac{3}{2}}^{-},\pm\frac{1}{2}>$
and $|P_{\mathrm{Bi},\frac{1}{2}}^{-},\pm\frac{1}{2}>$.
Here $S$ and $P$ denote corresponding orbits consisting of the new eigenstates and
the superscripts $\pm$ represent the parities of corresponding eigenstates.

According to the analysis of irreducible representations and projected orbits,
the touching point of the top of valence bands and the bottom of conduction bands
(denoted as $\Gamma_{8}^{-}$) is mainly composed of
$|P_{\mathrm{Bi},\frac{3}{2}}^{-},\pm\frac{3}{2}>$ and $|P_{\mathrm{Bi},\frac{3}{2}}^{-},\pm\frac{1}{2}>$ basis.
We simplify the notation of these four basis as $|J,j_{z}>$ with $J=\frac{3}{2}$ and $j_{z}=\pm\frac{3}{2},\pm\frac{1}{2}$.
Take the time-reversal and $O_{h}$ point-group symmetries into consideration,
a 4 $\times$ 4 Luttinger Hamiltonian \cite{LuttingerH-1956} can exactly describe
the $\Gamma_{8}^{-}$ bands around the $\Gamma$ point
if we arrange the 4 basis in the order of
$|\frac{3}{2},\frac{3}{2}>$,$|\frac{3}{2},\frac{1}{2}>$,
$|\frac{3}{2},-\frac{1}{2}>$,$|\frac{3}{2},-\frac{3}{2}>$, with the Hamiltonian given by
\begin{equation}\label{eq:eps}
H_{\mathrm{Luttinger}}(\vec{k})=\alpha_{0} \vec{k}^{2} I +\alpha_{1}(\vec{k}\cdot\vec{J})^{2}+\alpha_{2}\sum_{i=1}^{3}k_{i}^{2}J_{i}^{2},
\end{equation}
where $J_{i}(i=1,2,3)$ are spin-$\frac{3}{2}$ matrices and $\alpha_{i}(i=0,1,2)$
are parameters characterizing the band structures.
These three parameters are determined as
$\alpha_{0}\approx 205.3 {\AA}^{2}\mathrm{eV} $,$\alpha_{1}\approx -83.1 {\AA}^{2}\mathrm{eV }$,$\alpha_{2}\approx -22.5 {\AA}^{2}\mathrm{eV} $
by fitting the first-principle band structures around the $\Gamma$ point.
With this Luttinger Hamiltonian,
we can nicely describe the unique parabolic dispersion near $\Gamma$ at the Fermi level as shown in Fig.~\ref{fig:fm-3m-band}(c),
which is quite different from the linear Dirac dispersions.

\section{THE EFFECT OF Different STRAIN ON THE HIGH-PRESSURE Fm$\bar{3}$m PHASE}
\subsection{UNIAXIAL STRAIN ON the Fm$\bar{3}$m PHASE}
\begin{figure}[th]
\begin{center}
\includegraphics[width=0.48\textwidth]{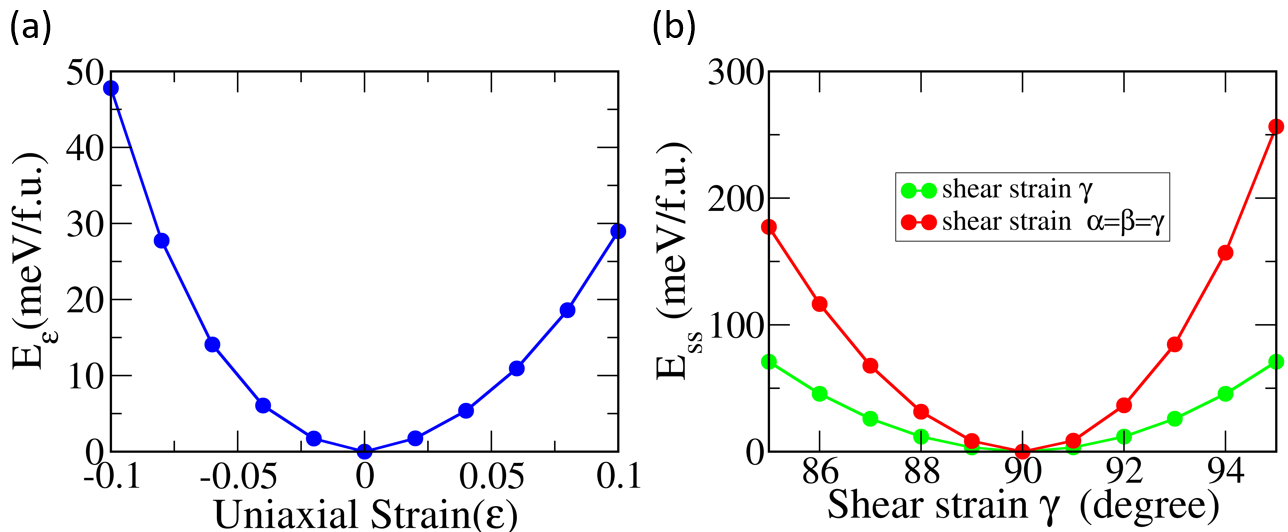}
\caption{%
(a) Strain energy of the Fm$\bar{3}$m Na$_{3}$Bi under
uniaxial strain with $\epsilon$ varying from -0.1 (compressive) to 0.1 (tensile).
(b) Shear strain energy along $\langle100\rangle$ (green) and $\langle111\rangle$ (red) axes
with the angular parameter $\gamma$ or $\alpha=\beta=\gamma$ ranging from 85$^{\circ}$ to 95$^{\circ}$, respectively.
}
\label{fig:fm-3m-shaear-strain-energy}
\end{center}
\end{figure}

When we impose a uniaxial strain $\epsilon $ along any orthogonal axis
on the original Fm$\bar{3}$m structure, it will change the symmetry of
the crystal from space group Fm$\bar{3}$m to I4/mmm.
Here, to keep the volume of the cell invariable,
when a tensile strain $\epsilon$ is applied along the $z$ axis,
we add at the same time a compressive strain $\frac{1}{\sqrt{1+\epsilon}}$ along the $x$ and $y$ axes, respectively.
It is obvious that this operation changes the symmetry
of the structure from Fm$\bar{3}$m to I4/mmm as well.
The uniaxial compressive or tensile strain energy
relative to the perfect Fm$\bar{3}$m structure
is shown in Fig.~\ref{fig:fm-3m-shaear-strain-energy} (a),
which indicates that the tensile strain somehow is easier
to obtain than the compressive one in this system.

With this uniaxial strain operation, as shown in Fig.~\ref{fig:fm-3m-linear} (a),
we can predict that the triply degenerate $p_{x,y,z}$ orbits without SOC at $\Gamma$ point
will split into a non-degenerated $p_{z}$
and a doubly degenerated $p_{x,y}$ orbits due to
the crystal-field splitting.

When SOC is considered, this uniaxial strain may
lead to the appearance of a Dirac crossing near the $\Gamma$ point.
In fact, the uniaxial strain changes the point group of the system
from O$_{h}$ to D$_{4h}$, which also affects
the $\mathbf{k}\cdot\mathbf{p}$ Hamiltonian dramatically.
As the permutation symmetry of $x, y, z$ directions is no longer preserved, 
$J_x$, $J_y$, $J_z$ is not convenient to be used as basis any more.
Here we use the following $\Gamma$ matrices:
\begin{equation}\label{eq:eps}
\begin{split}
&\Gamma^{1}=\frac{1}{\sqrt{3}}\{J_{y},J_{z}\},\Gamma^{2}=\frac{1}{\sqrt{3}}\{J_{z},J_{x}\},\Gamma^{3}=\frac{1}{\sqrt{3}}\{J_{x},J_{y}\}\\
&\Gamma^{4}=\frac{1}{\sqrt{3}}(J_{x}^{2}-J_{y}^{2}),\Gamma^{5}=J_{z}^{2}-\frac{5}{4},
\end{split}
\end{equation}
while the other ten $\Gamma$ matrices are given by
$\Gamma_{ab}=\frac{1}{2i}[\Gamma_{a},\Gamma_{b}]$.
The coexistence of TRS and IS constrains that
no $\Gamma_{ab}$ terms exist in the model Hamiltonian.
After a careful analysis of the symmetry and a tedious derivation,
we can give the character table of the $\Gamma$ matrices
and the polynomials of momentum $\vec{k}$ as shown in Table.~\ref{table:character-Fm-3m-uniaxial} using the same basis functions above-mentioned.

\begin{table}
\begin{ruledtabular}
\caption{The character table for the Fm$\bar{3}$m phase of Na$_{3}$Bi under uniaxial strain along z axis.}
   \begin{tabular}{l c c r}%
          \textbf{$\Gamma$}                      &    Representation                 &  T     & $\vec{k}$ \\
            $\Gamma_{0}=I$                       &    $\tilde{\Gamma}_{1}^{+}$       &  +     & $1,k_{x}^{2}+k_{y}^{2},k_{z}^{2}$ \\
            $\{\Gamma_{1},\Gamma_{2}\}$          &    $\tilde{\Gamma}_{5}^{+}$       &  +     & $\{k_{x}k_{z},k_{y}k_{z}\}$ \\
            $\Gamma_{3}$                         &    $\tilde{\Gamma}_{4}^{+}$       &  +     & $k_{x}k_{y}$ \\
            $\Gamma_{4}$                         &    $\tilde{\Gamma}_{3}^{+}$       &  +     & $k_{x}^{2}-k_{y}^{2}$ \\
            $\Gamma_{5}$                         &    $\tilde{\Gamma}_{1}^{+}$       &  +     & $1,k_{x}^{2}+k_{y}^{2},k_{z}^{2}$ \\
  \end{tabular}
  \label{table:character-Fm-3m-uniaxial}
\end{ruledtabular}
\end{table}

Finally, from Table.~\ref{table:character-Fm-3m-uniaxial}, our model Hamiltonian yields as
\begin{equation}\label{eq:eps}
\begin{split}
H&=\sum_{i=0}^{5}f_{i}(\vec{k})\Gamma_{i}=[a_{0}+b_{0}(k_{x}^{2}+k_{y}^{2})+c_{0}k_{z}^{2}]\Gamma_{0}\\
&+a_{12}(k_{y}k_{z}\Gamma_{1}+k_{x}k_{z}\Gamma_{2})+a_{3}k_{x}k_{y}\Gamma_{3}\\
 &+a_{4}(k_{x}^{2}-k_{y}^{2})\Gamma_{4}+[a_{5}+b_{5}(k_{x}^{2}+k_{y}^{2})+c_{5}k_{z}^{2}]\Gamma_{5}.
\end{split}
\end{equation}
The dispersion of above-mentioned model is
$E(\vec{k})=f_{0}(\vec{k})\pm \sqrt{f_{1}^{2}(\vec{k})+f_{2}^{2}(\vec{k})+f_{3}^{2}(\vec{k})+f_{4}^{2}(\vec{k})+f_{5}^{2}(\vec{k})}$ and both dispersions are doubly degenerated.
As a result, a band crossing of this model requires $f_{1}=f_{2}=f_{3}=f_{4}=f_{5}=0$, i.e. $k_{z}\neq 0\cap k_{x}=k_{y}=0 \cap a_{5}c_{5}<0$.
It means that we can always find a Dirac crossing along $k_{z}$ direction when $a_{5}c_{5}<0$ stands, and the location of the crossing is $\vec{k}=(0,0,\pm \sqrt{-\frac{a_{5}}{c_{5}}})$.
Otherwise, a gap near the $\Gamma$ point induced by $a_{5}\Gamma_{5}$ will always preserve.

When the bands are gapped, the existence of both the TRS and IS in the uniaxial-strained Fm$\bar{3}$m Na$_{3}$Bi
enables us to calculate $Z_{2}$ using Fu and Kane's method~\cite{TIs-with-inversion-2007-FuLiang}.
The results are listed in Table.~\ref{table:parity-Fm-3m-uniaxial}, which indicates Z$_{2}=(1, 000)$.

\begin{table}
\begin{ruledtabular}
\caption{The product of the parities for all the occupied bands at the eight TRIMs
for the Fm$\bar{3}$m phase of Na$_{3}$Bi under uniaxial strain along z axis.}
   \begin{tabular}{c c c c c c c c c c}%
          \textbf{TRIM}      & $\Gamma$  &  4N  &  2X  &  M  ;  & total\\
           Parity         &    --      &  +   &  +     &  +   ;  & --\\
  \end{tabular}
      \label{table:parity-Fm-3m-uniaxial}
\end{ruledtabular}
\end{table}

\begin{figure}[th]
\begin{center}
\includegraphics[width=0.45\textwidth]{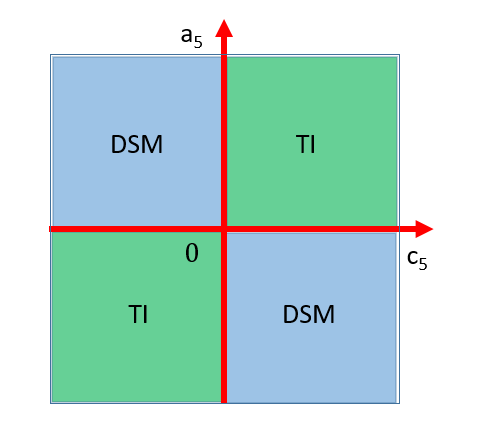}
\caption{%
Phase diagram of Fm$\bar{3}$m Na$_{3}$Bi under uniaxial strain
and shear strain along the $\langle111\rangle$ direction
from model Hamiltonian analysis using Equation (4) and (6), respectively.}
\label{fig:phasediagram}
\end{center}
\end{figure}

\begin{figure*}[th]
\begin{center}
\includegraphics[width=0.95\textwidth]{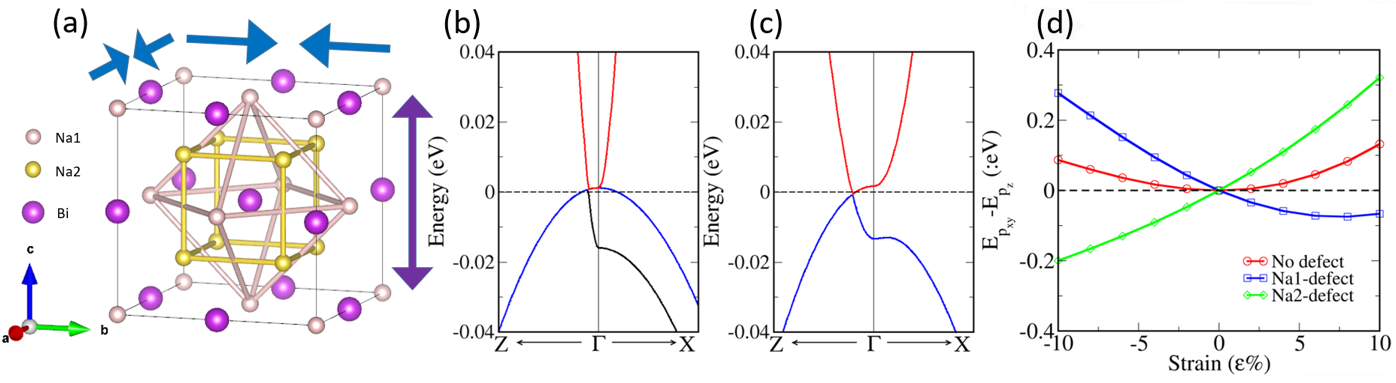}
\caption{%
  (a)schematic diagram of the Fm$\bar{3}$m phase with uniaxial strain.
Six Na1 atoms (pink) next-nearest neighbour around the body-centered Bi (purple) form an octahedron
while eight Na2 (yellow) nearest neighbour around form a cubic;
(b) and (c) The band structures near $\Gamma$ point of Na$_{3}$Bi in the space group of $I4/mmm$
which comes from uniaxial strain along z axis on the Fm$\bar{3}$m phase without and with spin orbital coupling.
The red and blue lines correspond to CBM and VBM respectively.
(d) The energy difference between $|Bi,p_{x,y}>$ and $|Bi,p_{z}>$ near the fermi level
vs. uniaxial strains in the range of $-10\%\leq\epsilon\leq 10\%$ for the intact, Na1-absent and Na2-absent Na$_{3}$Bi.
}
\label{fig:fm-3m-linear}
\end{center}
\end{figure*}

Note that the Z$_{2}$ index remains unchanged when we tune any parameters in the above-mentioned model because
band inversion between $|\frac{3}{2},\pm\frac{3}{2}>$ and $|\frac{3}{2},\pm\frac{1}{2}>$ can not bring the parity inversion
(the parities of both the two doublets at $\Gamma$ are -1).
So we can give resulting phase diagram of Fm$\bar{3}$m Na$_{3}$Bi under uniaxial strain shown in Fig.~\ref{fig:phasediagram}.
Which shows that, the system belongs to TI when $a_{5}c_{5}>0$, while it transforms into DSM in case of $a_{5}c_{5}<0$.
%

First-principles calculations indicates the Dirac crossing always exists under a strain $\epsilon$ ranging from $-10\%$ to $10\%$.
Which means $a_{5}c_{5}<0$ always stands for uniaxial-strained structure under strains within this range.



We have chosen a uniaxial strain of $\epsilon=-4\%$
onto the Fm$\bar{3}$m phase to verify our predictions,
the corresponding band structures without and with SOC are
showed in Fig.~\ref{fig:fm-3m-linear} (b) and (c) respectively.
As showed in Fig.~\ref{fig:fm-3m-linear}(c),
the appearance of this Dirac dispersion is very similar to
HgTe-class materials~\cite{WeylSM-HgTe-class-Ruan-NC-2016}, which can also be described
with an additional ${\Gamma}_5$ term in the Luttinger Hamiltonian ~\cite{LuttingerH-1956}.
The difference is,
in HgTe-class, with tensile stain along the $z$ axis,
the energy of $p_{z}$ becomes smaller than $p_{x,y}$,
while with compressive strain along the $z$ axis,
the energy of $p_{z}$ becomes larger than $p_{x,y}$
when SOC is ignored.
But in Na$_{3}$Bi, as shown in Fig.~\ref{fig:fm-3m-linear}(d),
both tensile and compressive stain leads to the same result
that $E_{p_{x,y}}$ is always larger than $E_{p_{z}}$.

Here we give a qualitative explanation.
Uniaxial strain along the $z$ axis generates a perturbation
$\mathcal{H}_{\mathrm{strain}:\mathrm{Na1;Na2}}=-g_{1,2}(J_{z}^{2}-\frac{5}{4})$.
Here $\mathcal{H}_{\mathrm{strain:Na1;Na2}}$ denotes the crystal perturbation on $|\mathrm{Bi},p>$
from the effect of the strain on Na1 and Na2, respectively.
From the unstrained structure in Fig.~\ref{fig:fm-3m-linear}(a),
we find that six Na1 atoms around
the body-centered Bi atom form an octahedron
while eight Na2 atoms form a cubic.
We further consider
the effect of the Na1-octahedron and Na2-cubic
on $|\mathrm{Bi},p>$ by removing all the Na2 or Na1 atoms, respectively.
As the blue and the green lines shown in Fig.~\ref{fig:fm-3m-linear}(d),
first-principle calculations of these two different cases indicate that
with the absence of Na2-cubic, $g_{1} < 0$ ($g_{1} > 0$)
corresponds to the tensile strain (the compressive strain);
while we remove Na2 cubic, $g_{2} > 0$ ($g_{2} < 0$)
corresponds the tensile strain (the compressive strain).
However, $g_{1} + g_{2} $ is always less than 0 for both tensile and compressive strain,
which means Na1-octahedron effects $Bi,p_{x,y}$ more in the tensile strain case,
while Na2-cubic effects more in the compressed strain case.

\subsection{SHEAR STRAIN ALONG the $\langle100\rangle$ DIRECTION ON the Fm$\bar{3}$m PHASE}
\begin{figure}[th]
\begin{center}
\includegraphics[width=0.48\textwidth]{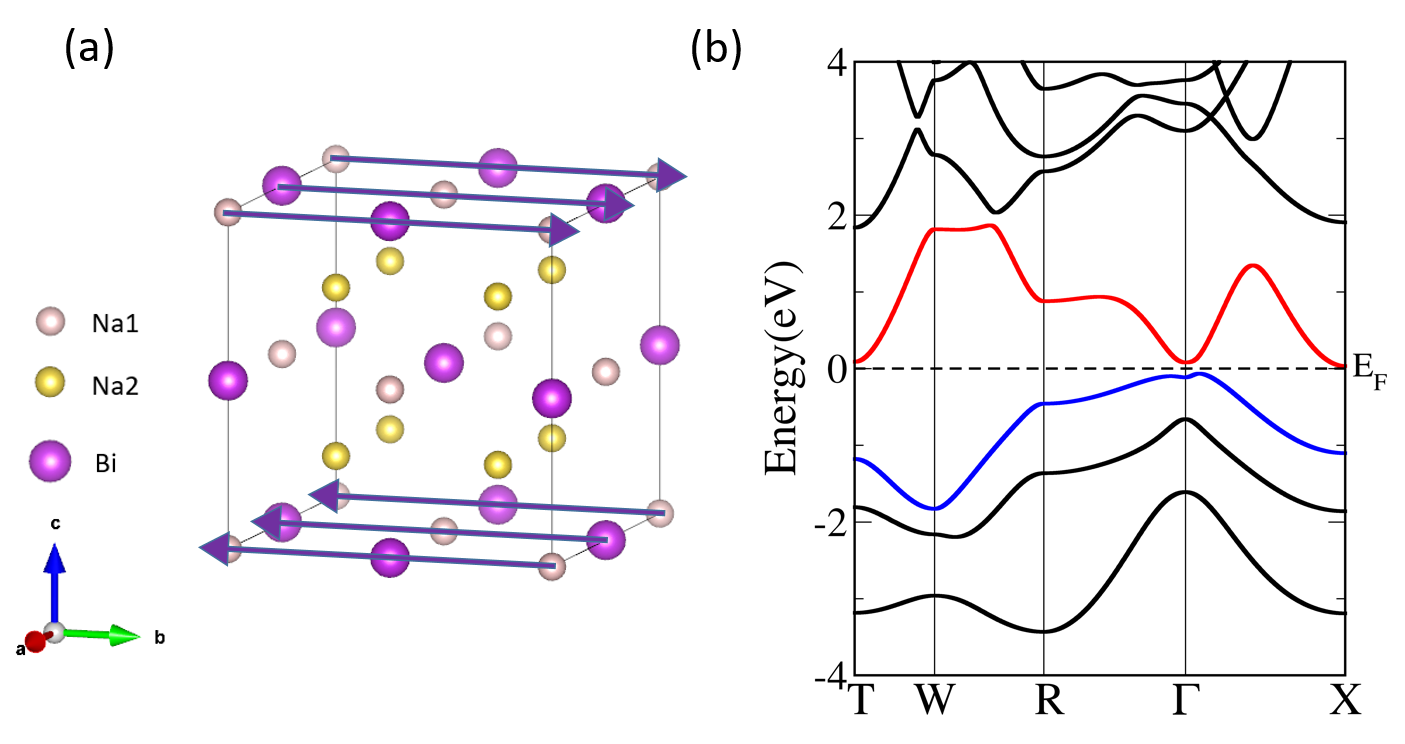}
\caption{%
(a)Crystal structure of Fm$\bar{3}$m Na$_{3}$Bi under $\langle100\rangle$ shear strain.
The purple arrows represent
shear strains on the top and bottom surfaces.
(b)Corresponding band structures (with SOC) of Na$_{3}$Bi
under the $\langle100\rangle$ shear strain.
The red and blue lines correspond to CBM and VBM, respectively.
}
\label{fig:fm-3m-shaear100}
\end{center}
\end{figure}

Shear strain might have different effect on the electronic
structures compared with tensile or compressive strain.
If we impose a shear strain along $\langle100\rangle$ direction
on the original Fm$\bar{3}$m structure,
we will get a structure in the space group of I/mmm
which belongs to the D$_{2h}$ point group.
In fact, as shown in Fig.~\ref{fig:fm-3m-shaear100} (a),
this operation only slightly changes one of the three lattice angles.
The green line in Fig.~\ref{fig:fm-3m-shaear-strain-energy} (b) shows
how shear strain energy E$_{ss\langle001\rangle}$ varies with
the angle parameter $\gamma$ ranging from 85$^{\circ}$ to 95$^{\circ}$.
It seems that the shear stain costs a large mount of energy.
Here we take the change of $\gamma$ from 90\degree to 86\degree as an example.
Electronic band structure shown in Fig.~\ref{fig:fm-3m-shaear100}(b) indicates that
the resulting structure is an insulator with a small gap of around 17 meV.
The D$_{2h}$ point group
ensures the existence of the inversion symmetry in the shear-strained structure,
which enables us to calculate $Z_{2}$ by multiplying the parities for all the occupied Bloch states at the eight TRIMs
using Fu and Kane's method~\cite{TIs-with-inversion-2007-FuLiang}.
The results are listed in Table.~\ref{table:parity-Fm-3m-100}, which clearly show $Z_{2}=1$.
Thus shear strain along the $\langle100\rangle$ direction brings the system from a semimetal into a TI.

\begin{table}
\begin{ruledtabular}
\caption{The product of the parities for all the occupied bands at eight TRIMs
for the Fm$\bar{3}$m phase of Na$_{3}$Bi with shear strain along $\langle100\rangle$ axis.}
   \begin{tabular}{c c c c c c c c c c}%
          \textbf{TRIM}      & $\Gamma$  &  2S  &  2R  &  2T  &   X ;  & total\\
           Parity          &    --      &  --   &  --   &  +   &   + ;  & --\\
  \end{tabular}
    \label{table:parity-Fm-3m-100}
\end{ruledtabular}
\end{table}

In fact, the shear strain changes the point group
from O$_{h}$ to D$_{2h}$, which also affects
the $\mathbf{k}\cdot\mathbf{p}$ Hamiltonian dramatically.
After a careful analysis of the symmetry and a tedious derivation,
one can give the character table of the $\Gamma$ matrice
and the polynomials of momentum $\vec{k}$ as shown in Table.~\ref{table:character-Fm-3m-100}.

\begin{table}
\begin{ruledtabular}
\caption{The character table for the Fm$\bar{3}$m phase of Na$_{3}$Bi with shear strain along $\langle100\rangle$ axis.}
   \begin{tabular}{l c c r}%
          \textbf{$\Gamma$}                      &    Representation                 &  T     & $\vec{k}$ \\
            $\Gamma_{0}=I$                       &    $\tilde{\Gamma}_{1}^{+}$       &  +     & $1,k_{x}^{2},k_{y}^{2},k_{z}^{2}$ \\
            $\Gamma_{1}$                         &    $\tilde{\Gamma}_{3}^{+}$       &  +     & $k_{y}k_{z}$ \\
            $\Gamma_{2}$                         &    $\tilde{\Gamma}_{2}^{+}$       &  +     & $k_{z}k_{x}$ \\
            $\Gamma_{3}$                         &    $\tilde{\Gamma}_{4}^{+}$       &  +     & $k_{x}k_{y}$ \\
            $\Gamma_{4}$                         &    $\tilde{\Gamma}_{1}^{+}$       &  +     & $1,k_{x}^{2},k_{y}^{2},k_{z}^{2}$\\
            $\Gamma_{5}$                         &    $\tilde{\Gamma}_{1}^{+}$       &  +     & $1,k_{x}^{2},k_{y}^{2},k_{z}^{2}$ \\
  \end{tabular}
  \label{table:character-Fm-3m-100}
\end{ruledtabular}
\end{table}

Finally, from Table.~\ref{table:character-Fm-3m-100}, our model Hamiltonian yields
\begin{equation}\label{eq:eps}
\begin{split}
H&=\sum_{i=0}^{5}f_{i}(\vec{k})\Gamma_{i}\\
 &=\sum_{i=0,4,5}(a_{i}+b_{ix}k_{x}^{2}+b_{iy}k_{y}^{2}+b_{iz}k_{z}^{2})\Gamma_{i}\\
 &+c_{1}k_{y}k_{z}\Gamma_{1}+c_{2}k_{z}k_{x}\Gamma_{2}+c_{3}k_{x}k_{y}\Gamma_{3}.
\end{split}
\end{equation}
It's clear that the band gap in the sheared structure
comes from $a_{4}\Gamma_{4}+a_{5}\Gamma_{5}$,
which depends on the point group D$_{2h}$ completely.
Similarly, a band crossing of above-mentioned model requires $f_{1}=f_{2}=f_{3}=f_{4}=f_{5}=0$,
i.e. $k_{\alpha}=k_{\beta}=0 \cap k_{\gamma} \neq 0\cap \frac{a_{4\gamma}}{b_{4\gamma}}= \frac{a_{5\gamma}}{b_{5\gamma}}<0$.
Here $\alpha,\beta,\gamma$ is a permutation of $\{x,y,z\}$.
It means that we can find a Dirac crossing along $k_{\gamma}$ direction only when $\frac{a_{4\gamma}}{b_{4\gamma}}= \frac{a_{5\gamma}}{b_{5\gamma}}<0$ stands,
and the location of the crossing is $k_{\alpha}=k_{\beta}=0, k_{\gamma}=\pm \sqrt{-\frac{a_{5\gamma}}{b_{5\gamma}}}$.
However, from our ab initio calculations, a gap always exists when shear deformation
along the $\langle100\rangle$ axis ($\gamma$) is in the range of $85-95$\degree.
In fact, $\frac{a_{4\gamma}}{b_{4\gamma}}= \frac{a_{5\gamma}}{b_{5\gamma}}<0$ is a very rigorous condition,
which can not be obtained without imposing other symmetries.
Thus a gap near $\Gamma$ will always occur.
On the other hand, the Z$_{2}$ invariant remains unchanged
because of the same reason discussed for the uniaxial-strained case.
Therefore, the splitting of crystal field together with SOC
generates $a_{4}\Gamma_{4}$ and $a_{5}\Gamma_{5}$ in the Hamiltonian,
and results in the topological nontrivial band gap at $\Gamma$ point.

\subsection{SHEAR STRAIN ALONG the $\langle111\rangle$ DIRECTION ON the Fm$\bar{3}$m PHASE}
\begin{figure}[th]
\begin{center}
\includegraphics[width=0.48\textwidth]{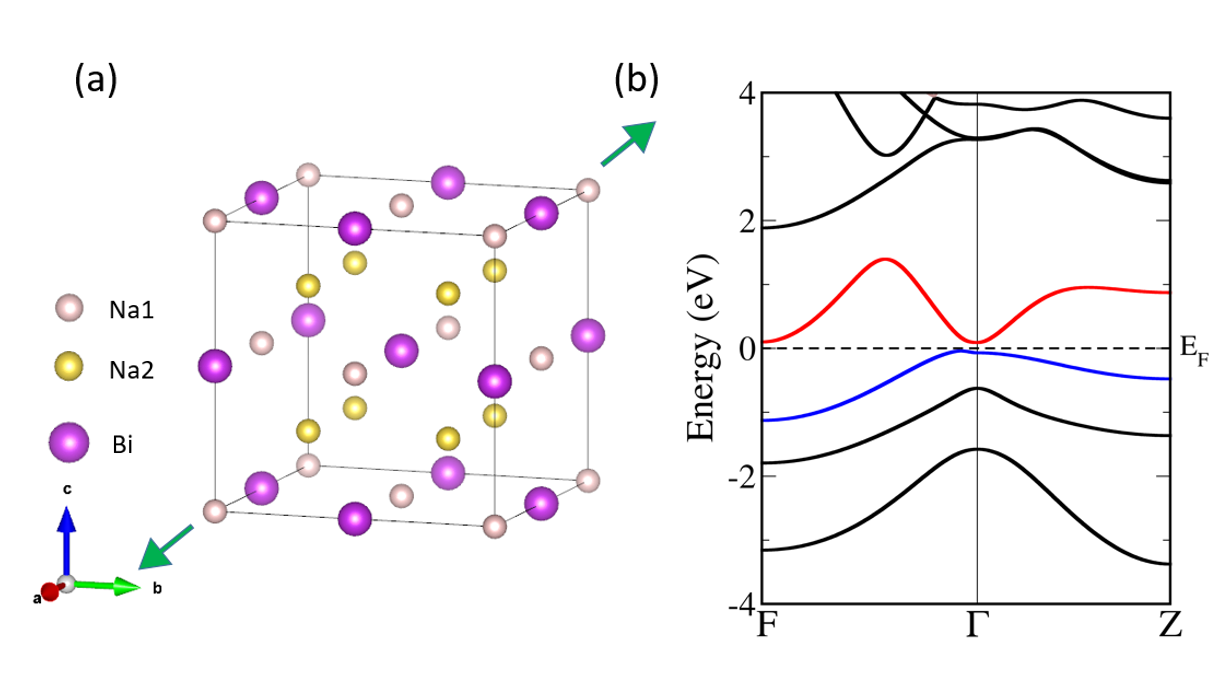}
\caption{%
(a)Crystal structure of Na$_{3}$Bi with Fm$\bar{3}$m symmetry
under shear strain in $\langle111\rangle$ direction.
The green arrows represents the shear direction.
(b)Corresponding band structure (with SOC) of Na$_{3}$Bi with
the $\langle111\rangle$ shear strain. The red and blue lines correspond to CBM and VBM respectively.
}
\label{fig:fm-3m-shaear111}
\end{center}
\end{figure}

Then we consider the case of shear strain
along the body diagonal direction as Fig.~\ref{fig:fm-3m-shaear111}(a) shows.
This type of strain turns the space group of the structure
from Fm$\bar{3}$m to R$\bar{3}$m (belongs to the D$_{3d}$ point group).
In fact, this type of strain can also be obtained by
changing the $\alpha ,\beta$ and $\gamma$ by the same amplitude,
which can be seen as a combination of shear strains
along the $\langle100\rangle$ $\langle010\rangle$ ,$\langle001\rangle$ directions.
The red line in Fig.~\ref{fig:fm-3m-shaear-strain-energy} (b) shows
how shear strain energy E$_{ss\langle111\rangle}$ varies with
the angle parameter $\alpha=\beta=\gamma$ ranging from 85$^{\circ}$ to 95$^{\circ}$.
Here we choose $\alpha =\beta =\gamma=92$\degree as an example.

The corresponding band structure shown in Fig.~\ref{fig:fm-3m-shaear111}(b) indicates an insulator phase.
Due to the preservation of IS, we calculate
the product of the parities for the occupied bands at all the eight TRIMs,
the result is shown in Table.~\ref{table:parity-Fm-3m-111}.
Parity inversion at the $\Gamma$ point leads to a nontrivial band topology with $Z_{2}=(1, 000)$,
which indicates that this $\langle111\rangle$ sheared structure is a strong TI.

\begin{table}
\begin{ruledtabular}
\caption{The product of the parities for all the occupied bands at the eight TRIMs
for the Fm$\bar{3}$m phase of Na$_{3}$Bi with shear strain along the $\langle111\rangle$ axis.}
   \begin{tabular}{c c c c c c c c c c}%
          \textbf{TRIM}      & $\Gamma$  &  3L  &  3FB  &  T  ;  & total\\
           Parity         &    --      &  +   &  +     &  +   ;  & --\\
  \end{tabular}
      \label{table:parity-Fm-3m-111}
\end{ruledtabular}
\end{table}

After similar analysis of the symmetry and derivation,
we can give the character table of $\Gamma$ matrices and the polynomials of momentum $\vec{k}$
for the case of $\langle111\rangle$ shear strain as Table.~\ref{table:character-Fm-3m-111} shows.
As a result, we can obtain the corresponding $\mathbf{k}\cdot\mathbf{p}$ Hamiltonian as
\begin{equation}\label{eq:eps}
\begin{split}
H&=\sum_{i=0}^{5}f_{i}(\vec{k})\Gamma_{i}\\
 &=[a_{0}+b_{0}(k_{x}^{2}+k_{y}^{2})+c_{0}k_{z}^{2}]\Gamma_{0} +a_{12}(k_{y}k_{z}\Gamma_{1}+k_{x}k_{z}\Gamma_{2})\\
 &+b_{12}[(-2k_{x}k_{y})\Gamma_{1}+(k_{x}^{2}-k_{y}^{2})\Gamma_{2}]\\
 &+a_{34}[2k_{x}k_{y}\Gamma_{3}+(k_{x}^{2}-k_{y}^{2})\Gamma_{4}]\\
 &+b_{34}[k_{y}k_{z}\Gamma_{3}-k_{x}k_{z}\Gamma_{4}]\\
 &+[a_{5}+b_{5}(k_{x}^{2}+k_{y}^{2})+c_{5}k_{z}^{2}]\Gamma_{5}.
\end{split}
\end{equation}
Similarly, a band crossing of this model requires $f_{1}=f_{2}=f_{3}=f_{4}=f_{5}=0$. It can be classified into two cases.
(a)$k_{x}=k_{y}=0 \cap k_{z} \neq 0\cap a_{5}c_{5}<0$,
(b)$\frac{a_{12}}{b_{12}}=-\frac{b_{34}}{a_{34}} \cap k_{z} \neq 0\cap (a_{5}b_{5}<0 \cup a_{5}c_{5}<0)$.
In the case of (b), if $\frac{-a_{5}a_{12}^{2}}{b_{5}a_{12}^{2}+c_{5}b_{12}^{2}}>0$, we can define $l=\sqrt{\frac{-a_{5}a_{12}^{2}}{b_{5}a_{12}^{2}+c_{5}b_{12}^{2}}}$,
then six Dirac points locating at $(\mp l,0,\pm\frac{b_{12}}{a_{12}}l)$, $(\pm\frac{1}{2}l,\pm\frac{\sqrt{3}}{2}l,\pm\frac{b_{12}}{a_{12}}l)$ and $(\pm\frac{1}{2}l,\mp\frac{\sqrt{3}}{2}l,\pm\frac{b_{12}}{a_{12}}l)$ can be found.
It's obvious that they are related to each other by $R_{3z}$ symmetries.
However, it should be noted that the condition of $\frac{a_{12}}{b_{12}}=-\frac{b_{34}}{a_{34}}$ 
in case (b) is very rigorous and can not be obtained without other symmetries,
i.e., case (a) is a unique condition for a stable DSM.
Thus, as discussed in the case of uniaxial strain, 
this system belongs to TI when $a_{5}c_{5}>0$, 
while it transforms into a DSM only in the case of $a_{5}c_{5}<0$.
As a result, this system have the same phase diagram as shown in Fig.~\ref{fig:phasediagram}.
Similar to the shear strain along $\langle100\rangle$ direction case, 
from our ab initio calculations, a gap always exists when shear deformation
along the the $\langle111\rangle$ axis ($\alpha =\beta =\gamma$) is in the range of $85-95$\degree.
Which indicates that $a_{5}c_{5}>0$ always stands for this shear strain.
Due to the similar reason as discussed in the case of uniaxial strain,
the gap induced by $a_{5}\Gamma_{5}$ term from D$_{3d}$ symmetry
will always remain and the Z$_{2}$ will not change.

\begin{table}
\begin{ruledtabular}
\caption{The character table for the Fm$\bar{3}$m phase of Na$_{3}$Bi
with shear strain along the $\langle111\rangle$ axis.}
   \begin{tabular}{l c c r}%
          \textbf{$\Gamma$}                      &    Representation                 &  T     & $\vec{k}$ \\
            $\Gamma_{0}=I$                       &    $\tilde{\Gamma}_{1}^{+}$       &  +     & $1,k_{x}^{2}+k_{y}^{2},k_{z}^{2}$ \\
            $\{\Gamma_{1},\Gamma_{2}\}$          &    $\tilde{\Gamma}_{3}^{+}$       &  +     & $\{k_{y}k_{z},k_{x}k_{z}\};\{-2k_{x}k_{y},k_{x}^{2}-k_{y}^{2}\}$ \\
            $\{\Gamma_{3},\Gamma_{4}\}$          &    $\tilde{\Gamma}_{3}^{+}$       &  +     & $\{2k_{x}k_{y},k_{x}^{2}-k_{y}^{2}\};\{k_{y}k_{z},-k_{x}k_{z}\}$ \\
            $\Gamma_{5}$                         &    $\tilde{\Gamma}_{1}^{+}$       &  +     & $1,k_{x}^{2}+k_{y}^{2},k_{z}^{2}$ \\
  \end{tabular}
\label{table:character-Fm-3m-111}
\end{ruledtabular}
\end{table}

\section{ SURFACE STATES OF THE Fm$\bar{3}$m PHASE WITHOUT and with STRAIN}
\begin{figure}[th]
\begin{center}
\includegraphics[width=0.48\textwidth]{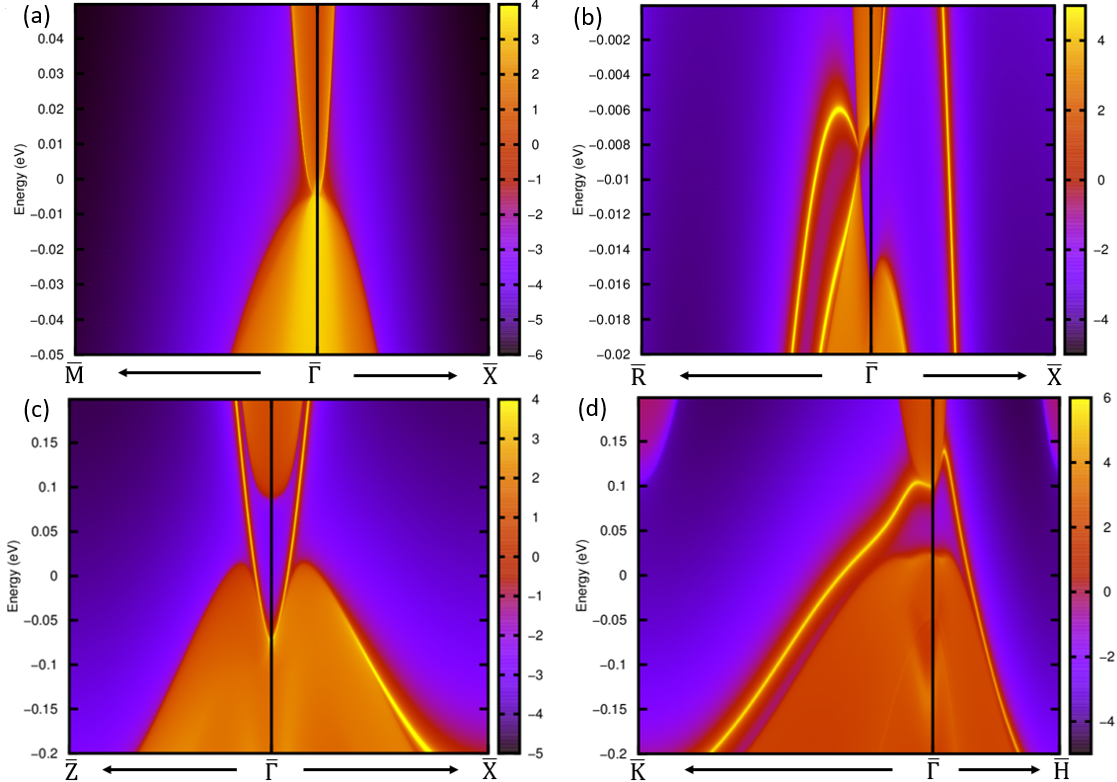}
\caption{%
The projected surface states of Fm$\bar{3}$m Na$_{3}$Bi
(a)without strain and terminated in $\langle100\rangle$ direction;
(b)under uniaxial strain and terminated in $\langle100\rangle$ direction;
(c)under shear strain in $\langle001\rangle$ direction and terminated in $\langle100\rangle$ direction;
(d)under shear strain in $\langle111\rangle$ direction and terminated in $\langle100\rangle$ direction
respectively.
}
\label{fig:ss}
\end{center}
\end{figure}

Exotic topological surface states is an important property
to identify various topological phases.
Based on the tight-binding model constructed with
MLWFs (maximally localised Wannier functions) method~\cite{mlwfs-1997,mlwfs-2001,MLWFs-RMPhys},
we have calculated the projected surface states of
the Fm$\bar{3}$m Na$_{3}$Bi without strain and with different type of strains,
as shown in Fig.~\ref{fig:ss}.
As showed in Fig.~\ref{fig:ss}(a), no topological protected
surface states can be found easily in the Fm$\bar{3}$m Na$_{3}$Bi without strain.
When we impose a uniaxial strain on Fm$\bar{3}$m Na$_{3}$Bi,
a Dirac crossing appears near the $\Gamma$ point in the bulk band structure.
And the corresponding non-trivial surface states
connecting the Dirac point also emerges, as showed in Fig.~\ref{fig:ss}(b).
As dicusssed above and showed in Fig.~\ref{fig:ss}(c) and Fig.~\ref{fig:ss}(d),
shear strains along the $\langle100\rangle$ and $\langle111\rangle$ directions
induce the original system into TIs, nontrivial metallic surface states can be found
in the gap.

\section{ CONCLUSION }
In conclusion, with first-principle calculations we investigate
the effect of stress/strain on Na$_{3}$Bi, a native 3D Dirac semimetal,
and find strains have large effects on the topological
band structures of this system.
We apply a uniaxial strain to break the R$_{3z}$ symmetry on the ambient P$\overline{3}$c1 phase
and find that this strain tunes Na$_{3}$Bi into a TI with a topological nontrivial gap at $\Gamma$ point.
Ab initio calculations show that the high pressure Fm$\bar{3}$m phase
is a new type of semimetal with the unique parabolic touching point
at $\Gamma$ point on the Fermi level,
which can be well described by a Luttinger Hamiltonian.
According to our calculations, uniaxial strain along the $\langle001\rangle$ direction can tune
the high pressure Fm$\bar{3}$m Na$_3$Bi from the parabolic semimetal into a DSM,
while shear strain along both the $\langle100\rangle$ and $\langle111\rangle$ directions
can tune the high pressure Fm$\bar{3}$m phase from the parabolic semimetal into a TI.
To gain more insights on these quantum phase transition from strain,
we derive three $\mathbf{k}\cdot\mathbf{p}$ models for the Fm$\overline{3}$m phase
and with all kinds of shear strains.
It is obvious that SOC together with the splitting of crystal field
from strains we imposed play key roles for the topological phase transitions in Na$_{3}$Bi.
In the end, we calculated surface states of Fm$\bar{3}$m Na$_3$Bi without strain and with different types of strains
to verify these topological transitions.
Different substrate might be used to introduce strain on
samples grew on them, which might be used to examine
the topological phase transitions studied in this work.

\section{ ACKNOWLEDGMENTS }
We thank the fruitful discussions with Huaiqiang Wang, Mengnan Chen, Feng Tang and Yongping Du.
This work is supported by the MOST of China (Grant Nos: 2016YFA0300404, 2015CB921202),
the National Natural Science Foundation of China (Grant Nos: 51372112, 11574133 and 11674165),
NSF Jiangsu province (No. BK20150012), 
the Science Challenge Project (No. TZ2016001),
the Fundamental Research Funds for the Central Universities (No. 020414380068/1-1),
Special Program for Applied Research on Super Computation of the NSFC-Guangdong Joint Fund (the second phase),
and Open Fund of Key Laboratory for Intelligent Nano Materials and Devices of the Ministry of Education (INMD-2016M01).
Part of the calculations were performed on the supercomputer in the HPCC of Nanjing University
and "Tianhe-2" at NSCC-Guangzhou.


%

\end{document}